\documentclass[sort, final, numberedheadings]{aipproc}
  
\layoutstyle{8x11single}

\begin{document}

%% commands for journal titles
\newcommand{\araa}{Ann. Rev. Astron. Astrophys.}
\newcommand{\aap}{Astron. Astrophys.}
\newcommand{\aj}{Astron. J.}
\newcommand{\apj}{Astrophys. J.}
\newcommand{\apjl}{Astrophys. J.}
\newcommand{\apjs}{Astrophys. J. Suppl.}
\newcommand{\mnras}{Mon. Not. Roy. Astron. Soc.}
\newcommand{\nat}{Nature}
\newcommand{\pasj}{Publ. Astron. Soc. Japan}

\title{Color bimodality: Implications for galaxy evolution}

\author{I. K. Baldry}{ address={Department of Physics \& Astronomy, 
  Johns Hopkins University, Baltimore, MD~21218, USA} }

\author{M. L. Balogh}{ address={Department of Physics, 
  University of Waterloo, N2L~3G1, Canada} }

\author{R. Bower}{ address={Department of Physics, 
  University of Durham, DH1 3LE, UK} }

\author{K. Glazebrook}{ address={Department of Physics \& Astronomy, 
  Johns Hopkins University, Baltimore, MD~21218, USA} }

\author{R. C. Nichol}{ address={Institute of Cosmology and Gravitation,
  University of Portsmouth, PO1~2EG, UK} }

\begin{abstract}
  We use a sample of 69726 galaxies from the SDSS to study the variation of
  the bimodal color-magnitude (CM) distribution with environment.  Dividing
  the galaxy population by environment ($\Sigma_5$) and luminosity
  ($-23<M_r<-17$), the $u-r$ color functions are modeled using double-Gaussian
  functions.  This enables a deconvolution of the CM distributions into two
  populations: red and blue sequences.  The changes with increasing
  environmental density can be separated into two effects: a large increase in
  the fraction of galaxies in the red distribution, and a small color shift in
  the CM relations of each distribution. The average color shifts are
  $0.05\pm0.01$ and $0.11\pm0.02$ for the red and blue distributions,
  respectively, over a factor of 100 in projected neighbor density.  The red
  fraction varies between about 0\% and 70\% for low-luminosity galaxies and
  between about 50\% and 90\% for high-luminosity galaxies.  This difference
  is also shown by the variation of the luminosity functions with environment.
  We demonstrate that the effects of environment and luminosity can be
  unified.  A combined quantity, $\Sigma_{\rm mod} = (\Sigma_5/{\rm Mpc}^{-2})
  \, + \, (L_r / L_{-20.2}) \,$, predicts the fraction of red galaxies, which
  may be related to the probability of transformation events. Our results are
  consistent with major interactions (mergers and/or harassment) causing
  galaxies to transform from the blue to the red distribution. We discuss this
  and other implications for galaxy evolution from earlier results and model
  the effect of slow transformations on the color functions. \footnote{Article
    written 2004 August 16th.  Refs.\ updated 2004 October 25th.  To appear
    in AIP Proc., The New Cosmology, eds.\ R.~E.\ Allen et al.}
\end{abstract}

\maketitle

\section{Introduction}
\label{sec-baldry:intro}

The study of galaxy properties in cosmology is important both for
understanding the formation and evolution of galaxies and for interpreting
measurements of large-scale structure.  For example, it is necessary to
determine the clustering bias of different types of galaxies relative to dark
matter in order to accurately quantify cosmological parameters. 

Galaxies were first classified based on their single-color morphological
properties by Hubble in the 1920's \citep{Hubble26}. The classification
followed a sequence of increasing complexity from ellipticals (E0-E7) to
lenticulars (S0) to spirals (Sa-Sc) to irregulars (Irr).\footnote{The
  adjectives ``early'' and ``late'' were used to describe the relative
  positions in the morphological sequence \citep{Hubble26}.  The direction
  from simple to complex forms was chosen from the precedent of stellar
  spectral types, where early types (O+B stars) generally have more simple
  spectral features than late types (K+M stars).  Ironically, the spectra of
  early-type galaxies are generally dominated by late-type stellar spectra and
  vice versa.} With the advent of color measurements, morphology-color
relations were established, with early types being generally redder in optical
colors than late types (for a review, see ref.~\citep{RH94}).

While there are many relationships between properties for various types of
galaxies, color and absolute magnitude are two of the most useful variables;
and unlike structural and spectral properties they are less dependent on
imaging resolution and aperture effects, respectively.  A color-magnitude (CM)
relationship for E+S0 galaxies was shown to have a shallow slope with a small
intrinsic scatter \citep{Faber73,VS77,APF81}.  This was related to a
metallicity-luminosity correlation, with more luminous galaxies having a
higher luminosity-weighted metallicity \citep{Faber73,Larson74,KA97}.  Spirals
also follow CM relationships but with larger intrinsic scatter
\citep{CR64,VG77,Griersmith80,Visvanathan81,TMA82}.  For spirals, the
luminosity correlations can be attributed to changes in star formation history
(SFH) \citep{PdG98}, dust attenuation \citep{tully98}, and/or metallicity
\citep{ZKH94}.

When all types are considered together, the color function of galaxies can be
approximated by the sum of two Gaussian functions that is a bimodal function
\citep{strateva01,baldry04}. This argues that the natural division in the
galaxy population is into two distributions; at least out to $z\sim1$
\citep{bell04red}.  Related spectral quantities such as from H$\alpha$
emission \citep{balogh04}, H$\delta$ absorption and the 4000\AA\ break
\citep{kauffmann03B}, and the derived star formation rate (SFR)
\citep{brinchmann04}, also produce a bimodal distribution. In
\S~\ref{sec-baldry:cm-distribution}, we review a quantitative study of the CM
distribution of galaxies, over all environments, using a two population model;
in \S~\ref{sec-baldry:enviro}, we analyze the environmental dependence of the
color bimodality; and in \S~\ref{sec-baldry:discussion}, we discuss the
results.

\section{The Overall Color-Magnitude Distribution}
\label{sec-baldry:cm-distribution}

\begin{figure}
\includegraphics[width=1.0\textwidth]{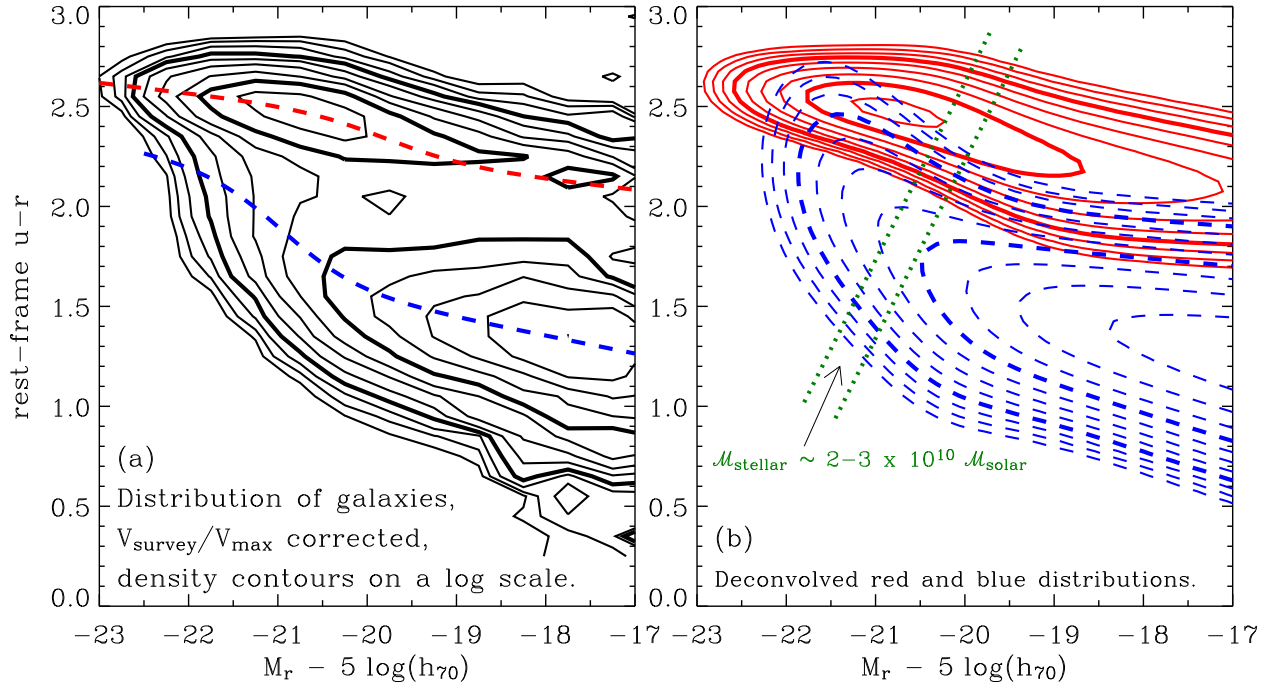}
\caption{Color-magnitude distributions.
  (\textbf{a}): Observed bimodal distribution, corrected for incompleteness.
  The contours are on a logarithmic scale in number density, doubling every
  two levels.  The dashed lines represent the color-magnitude relations of the
  red and blue sequences.  (\textbf{b}): Deconvolved and parameterized
  distributions.  The solid contours represent the red distribution and the
  dashed contours represent the blue distribution.  The dotted lines represent
  galaxies that have similar stellar masses, near the midpoints of the
  transitions.}
\label{fig-baldry:cm-trans}
\end{figure}

The Sloan Digital Sky Survey (SDSS)
\citep{york00,stoughton02,abazajian03,abazajian04} has dramatically improved
the statistics for studying CM relations of galaxies at low redshifts: with
over $10^5$ redshifts and associated five color photometry for $z<0.1$
galaxies.  \citet{baldry04} analyzed the distribution in color versus absolute
magnitude of a low-redshift sample of galaxies.
Figure~\ref{fig-baldry:cm-trans}(a) shows the CM distribution, corrected for
incompleteness, of all galaxies (isolated, in groups and in
clusters).\footnote{The cosmology assumed in this article is given by
  $(\Omega_{m},\Omega_{\Lambda})_0 = (0.3,0.7)$ with $H_0 = (h_{70})
  {\rm\,70\,km\,s^{-1}\,Mpc^{-1}}$. The data used is from the SDSS second data
  release \citep{abazajian04}, main galaxy sample \citep{strauss02}, with
  $0.01<z<0.08$. This provides a sample of 69726 galaxies with 99\% between
  $r$-band absolute magnitudes of $-23$ and $-17$. The magnitudes are
  $k$-corrected using a template fitting method \citep{blanton03kcorr}.} The
color $u-r$ was used as it spans the 4000\AA\ break and therefore is sensitive
to star formation history.  Even without a quantitative analysis, it is clear
that there are two dominant sequences which can be classically associated with
an E+S0 sequence (red distribution) and a spiral+irregular sequence (blue
distribution).

For each magnitude bin, the color function can be approximated as the sum of
two Gaussians, a bimodal function.  Two possible interpretations
for this are:
\begin{enumerate}
\item There is a continuous population but galaxies preferentially have
  certain colors, which depend on the luminosity. In this case, objects with
  intermediate colors are either transitioning or represent a middle sub-class
  of objects.
\item There are two separate populations that have associated properties with
  Gaussian-like color functions (normal distributions). Here, intermediate
  objects are not necessarily transitioning but instead belong to one of the
  two populations: with a probability of being in one or the other depending
  on the color (and/or other properties).
\end{enumerate}
In this article, we take the latter interpretation (which as shown later
provides a basis for illuminating how the CM distributions vary with
environment).  Thus, the distribution can be deconvolved into two dominant
components that have associated color-magnitude relations,
dispersion-magnitude relations and luminosity functions.  This is done by
fitting double-Gaussian functions to the color functions separated in absolute
magnitude bins. In addition, the mean and dispersion of the Gaussian are
constrained to vary smoothly with magnitude.  

This method differs from the classical approach of using cuts in morphology or
color to define classes and instead allows for a natural overlap.  The overlap
could arise from: photometric errors; degeneracy between dust reddening and
increasing stellar population age; stochastic variations in SFH (high past
average SFR with low recent SFR can be degenerate with the opposite case);
aperture effects for centrally-weighted colors\footnote{Because the $u$-band
  has low S/N in the SDSS data, the bimodality is better quantified using
  `model colors' \citep{stoughton02} that are derived by fitting
  de-Vaucouleurs or exponential profiles.  The profiles are defined using the
  $r$-band.} (bulge+disk can be degenerate with bulge only); complete
degeneracy between galaxies that had different formation mechanisms.

Figure~\ref{fig-baldry:cm-trans}(b) shows the deconvolved red and blue
distributions.  We describe below some of the results and points to note about
these sequences.  For full details, see ref.~\citep{baldry04}.
\begin{enumerate}
\item The color-magnitude relations are not well fit by straight lines.  A
  good fit is obtained with a straight line plus a tanh function, interpreted
  as a general trend plus a transition.
\item The color dispersion at the faint end of the red distribution is
  significantly higher than at the bright end.  This is consistent with the
  low-luminosity red distribution galaxies forming their stars later on
  average that the more luminous galaxies (e.g., fig.~1 of ref.~\citep{BKT98}
  shows that younger stellar populations produce a higher dispersion in CM
  relations). If this is the case, then the low-luminosity red sequence may
  not be in place at high redshift.  Confirming this, recent results for
  clusters at $z\sim1$ find a reduced number of galaxies in this part of the
  CM distribution \citep{delucia04,kodama04}.
\item The blue distribution gets significantly redder for galaxies more
  luminous than $M_r\approx-20$. This can be interpreted as being caused by
  the increasing importance of dust with increasing luminosity; and a
  reduction in the specific SFR for the most luminous blue-distribution
  galaxies (which can in fact be quite red).
\item The significant overlap around $M_r\sim-21.5$ and $u-r\sim2.4$ between
  the two distributions is at least partly due to a degeneracy between dust
  reddened late types and old stellar population early types. However, we note
  that the nature of the parametric fitting may overestimate the overlap.
\item If the galaxies' luminosities are converted to stellar masses, the
  transitions occur around the same mass as that found using spectroscopic
  measurements by \citet{kauffmann03B} ($3\times10^{10}{\cal M}_{\odot}$).
  This is an important confirmation of this transition mass using photometry,
  which uses apertures that scale with the size of the galaxy, as opposed to
  spectroscopy, which uses a $3''$ aperture (see also discussion by
  \citet{Kannappan04}).  The transition involves a change in the properties of
  both distributions and a change over in dominance from one to the other.
\item Other results suggest that galaxy environments affect the fraction of
  galaxies in each distribution but have little effect on the properties of
  galaxies within a distribution \citep{budavari03,balogh04,balogh04bimodal}
  (\S~\ref{sec-baldry:enviro}).
\end{enumerate}

What causes the bimodality?  Our results show that there must be two distinct
types of galaxies: a passively evolving red population and a separate
population of blue star-forming galaxies. The bimodality means that there
cannot be a continuous spread in galaxy properties, and suggests that galaxies
must move rapidly between the two populations.  The red distribution is
generally associated with (morphologically classified) early-type galaxies,
which have more virialized motions of stars and less dust. Simulations have
shown that major mergers can produce elliptical galaxies
\citep{TL79,Barnes88,BH92} and the gas or dust may be expelled by a burst of
star formation \citep{JW85}. Therefore, it seems reasonable that mergers (or
other major interactions) are the cause of the bimodality, with the red
distribution formed by violent means and the blue distribution from more
quiescent accretion. In the next section, we analyze the bimodality of the CM
distribution as a function of environment to test this idea further.

\section{Environmental Dependence}
\label{sec-baldry:enviro}

It has been known for some time that there is a higher proportion of
early-type galaxies in regions of high environmental density.
\citet{Dressler80} quantified a relationship between local galaxy density and
the fraction of E, S0 and spiral galaxies, with increasing E and S0
populations with increasing density.  To test how the color bimodality depends
on environment, we divided the galaxy population into five environmental
bins.

The environmental density was estimated using a surface density given by
$\Sigma_5 = 5 / (\pi r^2)$ where $r$ is the projected distance to the
fifth-nearest spectroscopically-confirmed neighbor (within $\pm
1000{\rm\,km\,s^{-1}}$) brighter than $M_r=-20$. This is a two dimensional
density given in units of ${\rm Mpc}^{-2}$.  Determining a true three
dimensional density is non-trivial because of peculiar velocities. In low
density regions, the conversion is approximately given by $\rho_5 = \Sigma_5 /
(28{\rm\,Mpc})$ because peculiar velocities will be small and the diameter of
the cylinder ($>8{\rm\,Mpc}$ for $\Sigma_5<0.1{\rm\,Mpc}^{-2}$) will be of
order the height ($28{\rm\,Mpc}$ at $z=0.05$). At higher densities, for most
galaxies the density will be increasingly underestimated using this conversion
because galaxies will be closer together than their velocities imply.  In
addition, galaxies may not be observed spectroscopically because of
fiber-placement restrictions and this restriction is more severe in high
density regions (though we note that $>92$\% of the target galaxies are
observed \citep{blanton03tile}).  Two orders of magnitude in $\Sigma$ may
correspond to about three orders of magnitude in $\rho$ (cf.\ fig.~4 of
ref.~\citep{Dressler80}).

In cases where the edge of the survey is closer than the fifth-nearest
neighbor, the distance to the boundary is used to determine an upper limit to
the density while the distance to the fifth-nearest neighbor is used for a
lower limit. The midpoint between the limits is used to determine which bin a
galaxy falls in and galaxies are {\em not rejected} if (i) both limits fall in
the least or most dense bin or (ii) the uncertainty in the density is less
than the width of the smallest bin. From the SDSS second data release, main
galaxy sample, 59085 are retained out of 69726 galaxies in the redshift range
0.01--0.08.

\begin{figure}
\includegraphics[width=1.0\textwidth]{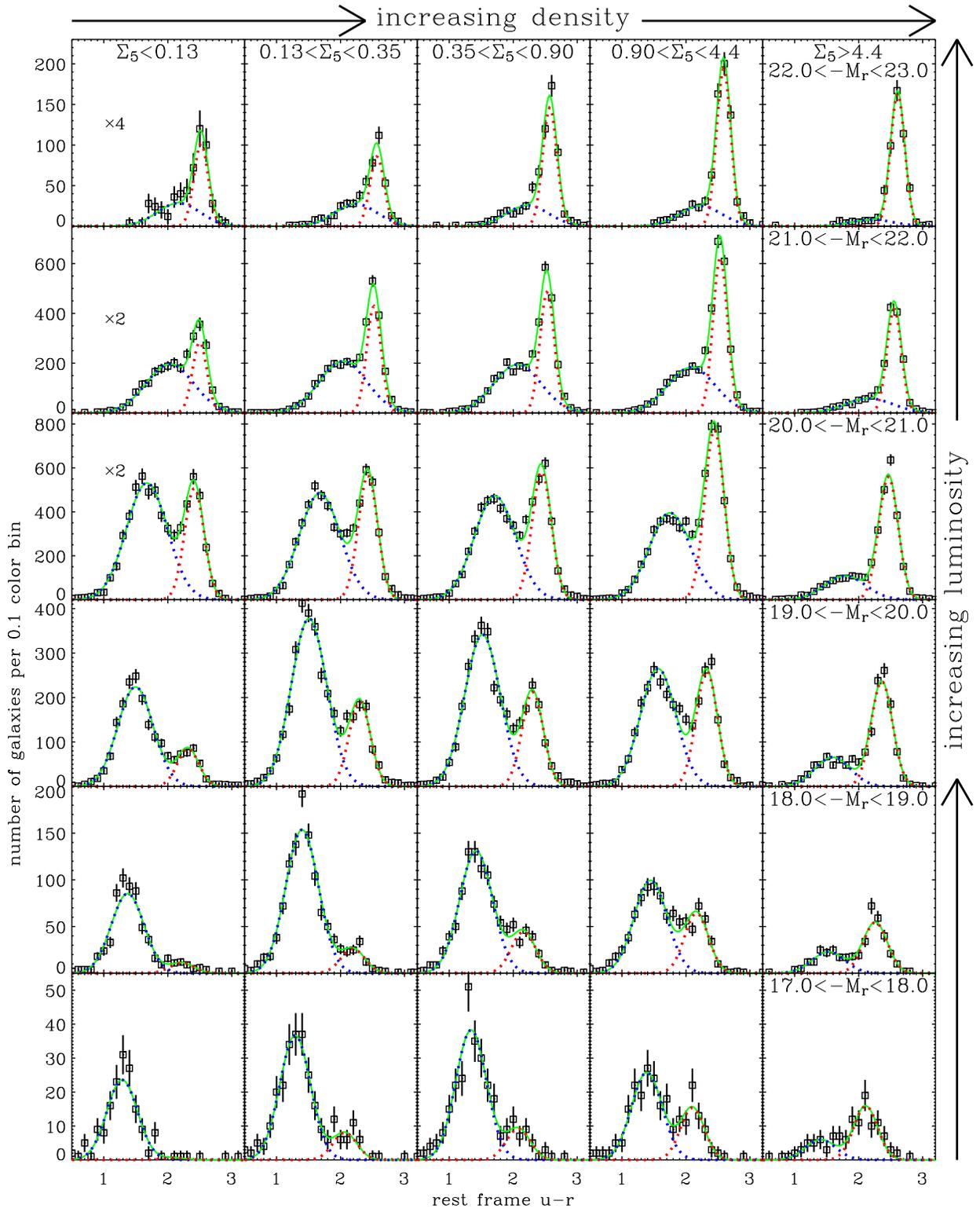}
\caption{Galaxy color functions
  in bins of environmental density ($\Sigma_5/{\rm Mpc}^{-2}$) and luminosity
  ($M_r$). The squares represent the data points with error bars, while the
  gray solid lines represent the double-Gaussian fits. The dotted lines
  represent the individual Gaussian functions. Plots where the counts have
  been scaled are marked $\times$2 or $\times$4.}
\label{fig-baldry:color-enviro}
\end{figure}

Figure~\ref{fig-baldry:color-enviro} shows the $u-r$ galaxy color functions in
bins of luminosity and projected density.\footnote{These results were
  determined using the SDSS second data release \citep{abazajian04}.  Earlier
  results presented by \citet{balogh04bimodal} used the first data release
  \citep{abazajian03}.  Also, here we use Petrosian magnitudes for the
  absolute magnitudes, whereas the earlier results used model magnitudes. Both
  use model colors. The variance used for the fitting is a modified Poisson
  noise estimate, given by $N+2+(0.05\,\bar{N})^2$, where $\bar{N}$ is the
  average counts over all 28 color bins (0.45--3.25). The $+2$ term allows for
  a more realistic estimate of the variance for low counts; and the 5\% factor
  allows for systematic errors and some deviation from Gaussian
  distributions.} The middle density bins have equal numbers of galaxies,
while the least and most dense bins have half as many. The data are fitted
with double Gaussians, with the mean and amplitude of each distribution varied
in all bins, while the dispersions are allowed to vary as a function of
luminosity only.  The data are well modeled by this parametric form. Thus,
there is no major distortion to either distribution as a function of
environment; and at fixed luminosity, the dominant change with environment is
the fraction of galaxies in each distribution. For further discussion, see
ref.~\citep{balogh04bimodal}.

\begin{figure}
\includegraphics[width=1.0\textwidth]{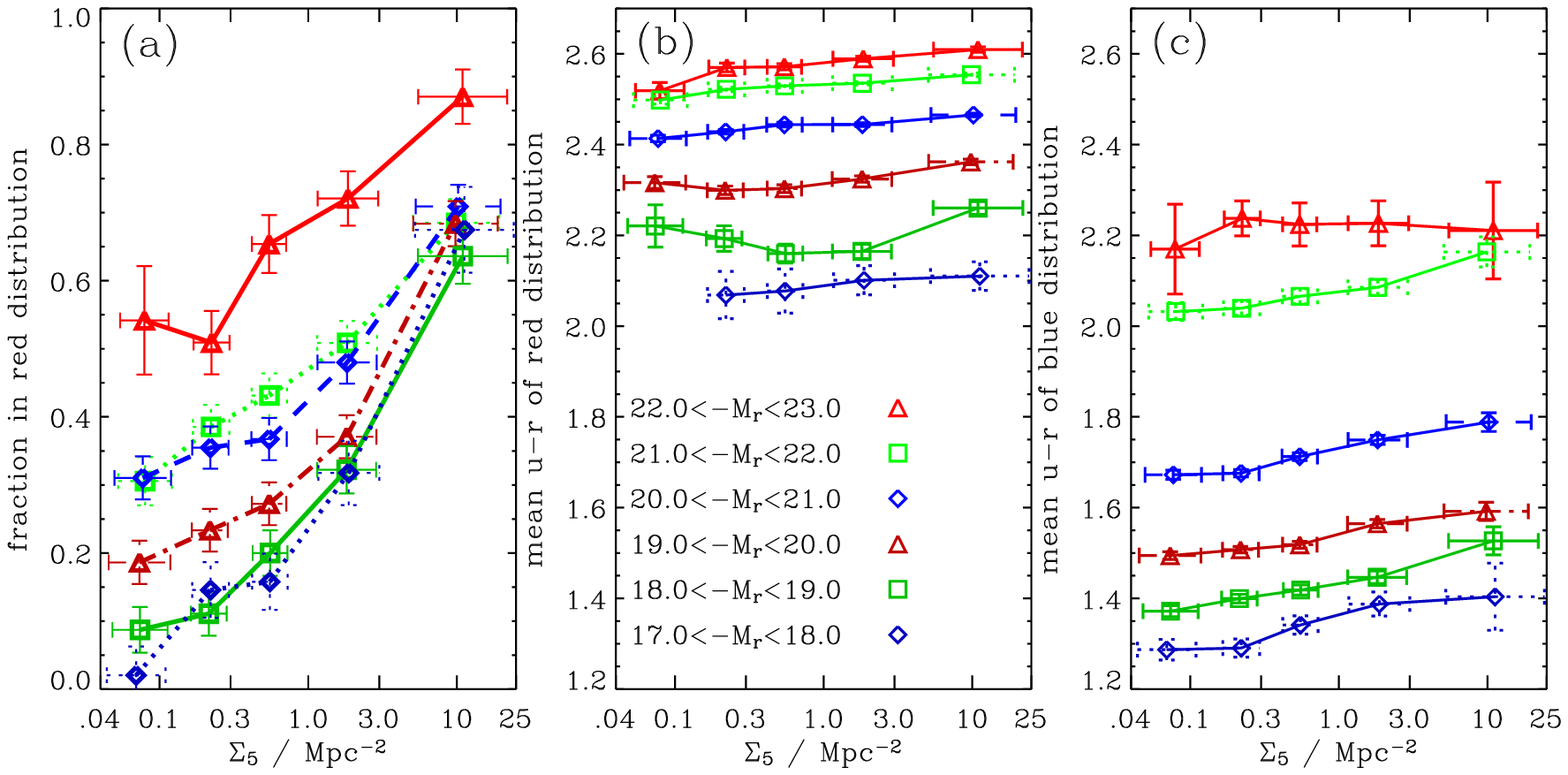}
\caption{The fraction of galaxies in the red distribution (\textbf{a}) 
  and the mean color of each distribution (\textbf{b}, \textbf{c}) as a
  function of environmental density. The lines and symbols represent different
  luminosity bins.  (\textbf{b}, \textbf{c}): The average change in the mean
  color over a factor of 100 in $\Sigma_5$ is $0.05\pm0.01$ and $0.11\pm0.02$
  for the red and blue distributions, respectively. These shifts were
  determined using weighted averages from straight-line fits over the six
  luminosity bins.}
\label{fig-baldry:frac-ampl-mean}
\end{figure}

Figure~\ref{fig-baldry:frac-ampl-mean} shows the change in the fraction in the
red distribution and the change in the mean color of each distribution versus
environment. While the fraction on the red sequence increases by $\sim50$\%,
the mean colors only increase by $0.05\pm0.01$ and $0.11\pm0.02$ for the red
and blue sequences, respectively, over a factor of 100 in projected density.
Thus, we can separate two effects of increasing environmental density, major
and minor: an increased trigger rate for transforming galaxies from the blue
to the red distribution; and a modest reddening of each distribution, which
could be caused by increasing average stellar age with increasing density.  If
the latter explanation were correct for the red distribution, it implies a
difference in luminosity-weighted stellar age of about 1\,Gyr between the
lowest and highest density environments (cf.\ 
refs.~\citep{bernardi03C,hogg04}).

The projected density has a continuous effect on the populations, at low
redshift, considering both the major and minor effects. This suggests that
cluster specific processes such as ram-pressure stripping \citep{FS80} do not
play a major role. Other work has also found that the primary effect is local
density rather than cluster dynamics \citep{PG84,balogh04bimodal,depropris04}.
For example, \citet{PG84} found that the morphology-density relation was
similar for galaxies in groups and in/around rich clusters.

\begin{figure}
\includegraphics[width=1.0\textwidth]{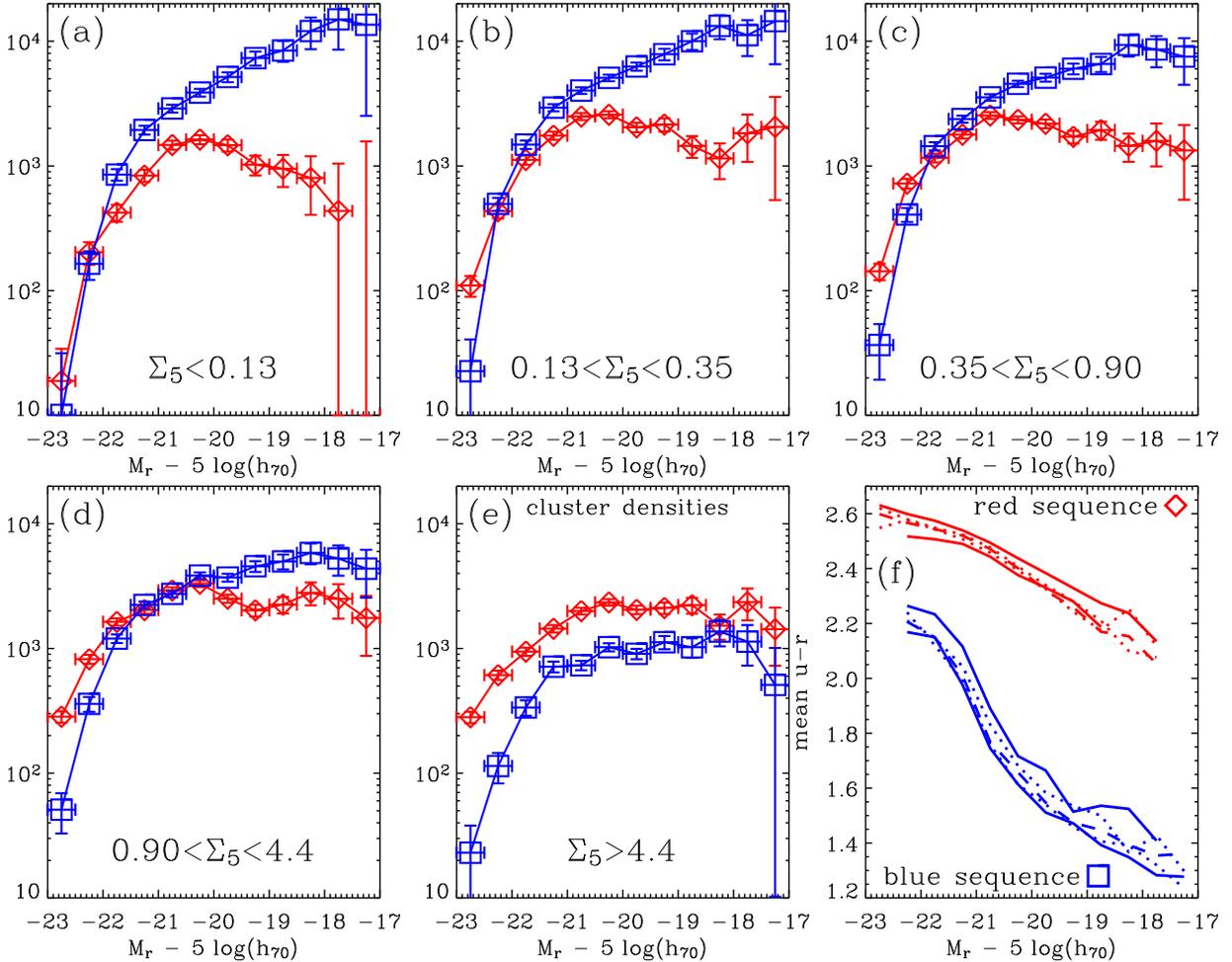}
\caption{(\textbf{a}--\textbf{e}): Luminosity functions 
  for different environmental densities.  The gray squares represent the blue
  sequence while the diamonds represent the red sequence.  The $y$-axis scale
  represents the completeness-corrected number per magnitude from the
  $0.01<z<0.08$ sample ($\sim10^7{\rm\,Mpc^3}$). (\textbf{f}):
  Color-magnitude relations for all densities.  The solid lines represent the
  lowest and highest density bins; the dashed line, the middle density bin;
  and the dotted lines, the remaining bins.  The sequences become slightly
  redder with increasing density (see also
  Fig.~\ref{fig-baldry:frac-ampl-mean}); and they are only plotted where the
  error in the $u-r$ value is less than 0.07 and 0.12 for the red and blue
  sequences, respectively (typical formal errors are 0.01--0.04).}
\label{fig-baldry:lum-funcs}
\end{figure}

An alternative view of the numbers in the red and blue sequences as a function
of environment is given by the luminosity functions.
Figure~\ref{fig-baldry:lum-funcs}(a-e) shows these for the five environmental
bins (with the highest density bin representing typical cluster densities).
Completeness corrections were computed taking into account the magnitudes,
redshifts and environmental densities of the galaxies. While the luminous
cutoffs remain similar for all bins, the faint-end slopes are changing.  In
the highest density bin, the two sequences have a similar slope, whereas in
the lowest density bin, the blue sequence has a steep slope and the red
sequence a shallow slope. In particular, there are few low-luminosity
red-distribution galaxies in the lowest density environmental bin, relative to
blue-distribution galaxies. Note that the data are only volume limited for
$M_r<-20$ and therefore the measured faint-end slopes depend significantly on
large-scale structure variations with redshift (using the $V_{\rm
  survey}/V_{\rm max}$ method); whereas, the relative faint-end slopes do not.

It is not possible to quantitatively compare these results with previous work
for two reasons: (i) this method involves a deconvolution, the double-Gaussian
fitting; and (ii) the definition of red/blue (early/late) types varies
slightly with density (the minor effect). Qualitatively, there is good
agreement with 2dFGRS \citep{colless01,colless03} results, in the sense that
the faint-end slopes for early- and late-type galaxies are similar in clusters
\citep{depropris03} but different in the field \citep{madgwick02} (see also
ref.~\citep{croton04}).  Here, the early/late division was based on a spectral
type.  There is disagreement with cluster LFs based on SDSS data using a cut
at $u-r=2.2$ to divide early/late \citep{goto02}, where the blue galaxies have
a significantly steeper slope.  This is because a color cut does not take
account of the CM relations of the red/blue sequences (see
Fig.~\ref{fig-baldry:color-enviro}, a cut at 2.2 slices the red distribution
in half at low luminosities).

\begin{figure}
\includegraphics[width=1.0\textwidth]{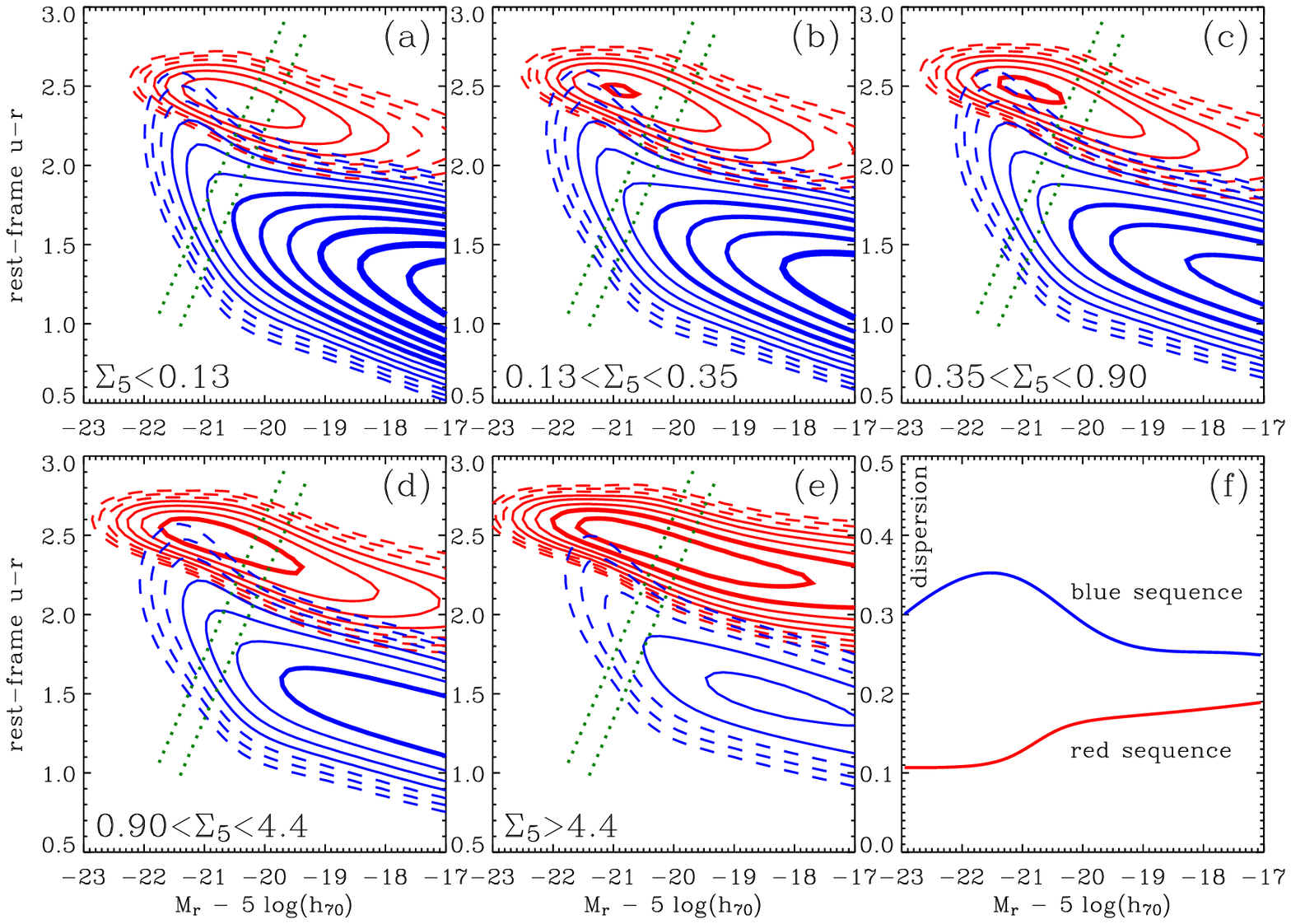}
\caption{(\textbf{a}--\textbf{e}): Parameterized reconstructions 
  of the color-magnitude distributions as a function of environmental density
  (normalized by total luminosity density).  The gray contours represent the
  blue sequence while the black contours represent the red sequence.  For this
  figure: the fitted means and dispersions have been smoothed using tanh
  functions plus a straight line or a quadratic; and the luminosity functions
  using single or double Schechter functions \citep{baldry04}. The dotted
  lines represent galaxies that have similar stellar masses (see
  Fig.~\ref{fig-baldry:cm-trans}).  (\textbf{f}): Dispersion versus magnitude
  used for each sequence derived from the best fit on the {\em assumption}
  that the dispersion does not vary with density. Note that the measured
  dispersion includes observational uncertainties. The changes in dispersion
  are not obvious in Panels (a-e) because changes in the contours are
  generally dominated by the effects of the luminosity functions.}
\label{fig-baldry:cm-contours}
\end{figure}

Figures~\ref{fig-baldry:lum-funcs}(f) and~\ref{fig-baldry:cm-contours}(a-e)
show the CM relations and distributions, respectively. The CM relations change
slope at similar magnitudes over all the environmental densities (for the red
and blue distributions, separately).  In particular, the dotted lines in
Fig.~\ref{fig-baldry:cm-contours} (representing similar stellar masses) cut
through the steepest part of all the CM relations.  This shows that the
transition mass \citep{kauffmann03B,baldry04} is similar in all environments.
Thus, this difference between low and high luminosity galaxies depends on the
mass of the galaxy and not the environment. \citet{DW03} suggest that this
division is related to supernova feedback.

The CM distributions in Fig.~\ref{fig-baldry:cm-contours} visually emphasize
the main points of this article: (i) that the major effect with increasing
density is the increase in the fraction of galaxies on the red sequence; and
(ii) that, by comparison, the CM distributions of each sequence vary in a
minor way, with a small shift toward redder colors with increasing density.
Intriguingly, if higher-density environments went through rapid evolution but
basically started from a similar population to low-density regions today, then
increasing density also represents increasing time. Notably both the luminous
and faint ends of the red sequence increase in number by more than the
$M_r\sim-20.5$ galaxies, by transformations from the blue sequence [cf.\ 
Fig.~\ref{fig-baldry:lum-funcs}(a,e)].\footnote{Studies looking at the mean
  environment \cite{hogg03} and clustering \citep{zehavi04} of galaxies as a
  function of color and magnitude (`the other side of the coin' to our
  analysis), see the same effect. Here, luminous and faint red galaxies are
  found in more dense environments than intermediate red galaxies.} The
luminous-red galaxies may have been formed by major mergers (also involving
red-sequence galaxies) but this is unlikely to be the cause for the faint-red
galaxies. This is because low-mass to low-mass galaxy mergers should be rarer
than simply low-mass galaxies being `cannibalized' by massive galaxies (minor
mergers). Instead, close encounters of some kind may be enough to cause the
transformation.

For our analysis, we have assumed that the color dispersions of each sequence
do not depend on environment [Fig.~\ref{fig-baldry:cm-contours}(f)].  The
best-fit dispersions were obtained by minimizing the combined $\chi^2$ over
all the environmental bins.  Allowing the dispersions to vary would
significantly increase the complexity of the double-Gaussian fitting; and the
results would rely more strongly on the assumption that the distributions are
exactly Gaussians. Visual inspection of Fig.~\ref{fig-baldry:color-enviro}
shows that assumption of dispersions varying with luminosity only is
reasonable.  There are only a couple of plots where a small change in
dispersion would clearly benefit the fit (e.g.\ bin with $M_r=-18.5$ and
$\Sigma_5<0.13$).  This is not to say that any change in color dispersion
would not be interesting (for constraining star-formation and merging
histories of these galaxies \citep*{BKT98}) but that it would require more
data (future SDSS data releases) and/or a suitable technique forcing the
dispersions to vary smoothly with environment and luminosity.

\subsection{Combining neighbor density and luminosity}

Figure~\ref{fig-baldry:frac-ampl-mean}(a) shows that there is some difference
between the luminosity bins, particular at the lowest densities. There is a
higher fraction of red-distribution galaxies in the most luminous bin compared
to the lower luminosity bins.  However, the density measure $\Sigma_5$ only
measures the number density of bright neighbors and does not account for the
luminosity of the galaxy in question, which could be regarded as the very
local density. In other words, there must be a local density peak to form a
high-mass galaxy even if that galaxy is isolated (at the present time).

\begin{figure}
\includegraphics[width=1.0\textwidth]{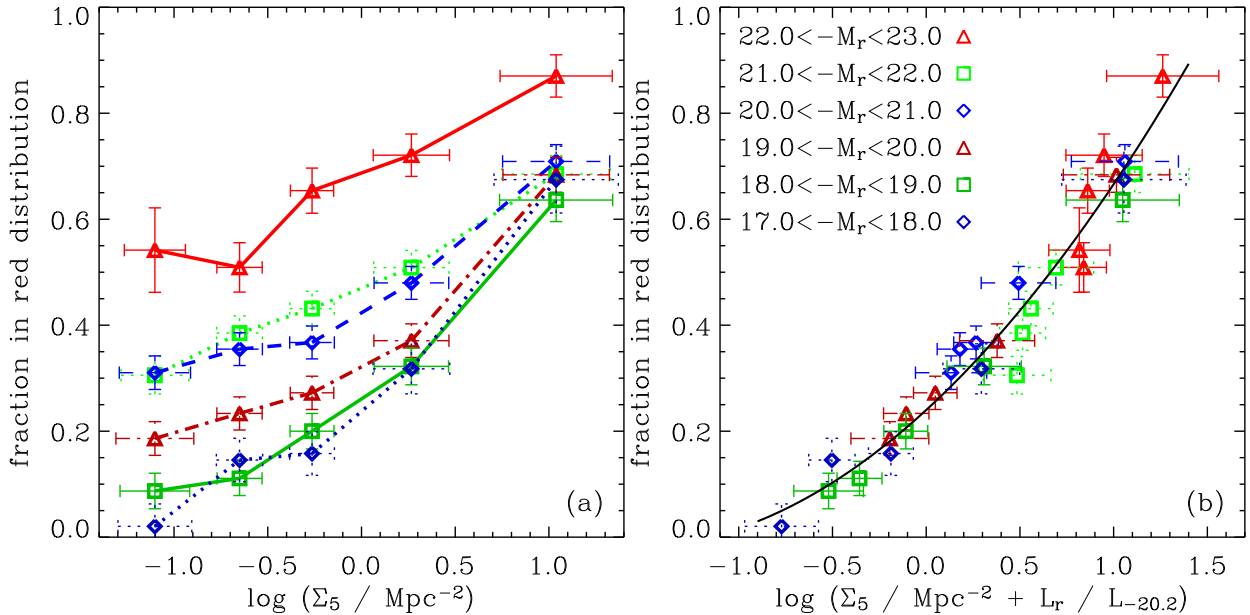}
\caption{The fraction of galaxies in the red distribution 
  as a function of environmental density. Panel~(\textbf{a}) uses the
  projected density $\Sigma_5$ [same as
  Fig.~\ref{fig-baldry:frac-ampl-mean}(a)] while Panel~(\textbf{b}) uses a
  combination of the projected density and the $r$-band luminosity (normalized
  by the luminosity for $M_r=-20.2$). (\textbf{b}): The solid line represents
  a fit to the data points using all luminosity bins.}
\label{fig-baldry:frac-ampl-adj}
\end{figure}

To unify the luminosity and the neighbor density, we summed the two values
using only one normalization parameter, which was adjusted so that the new
parameter was the optimal predictor for the fraction in the red distribution.
The combined quantity is given by
\begin{equation}
  \Sigma_{\rm mod} = (\Sigma_5/{\rm Mpc}^{-2}) \, + \, (L_r / L_{-20.2}) \: ,
\end{equation}
where $L_{-20.2}$ is the fitted normalization luminosity
($5.25\times10^{21}\,{\rm W\,Hz^{-1}}$).  This summation of the two terms is
highly suggestive of a combined probability from an environmental and a
host-galaxy mass term.  Figure~\ref{fig-baldry:frac-ampl-adj}(b) shows the
relationship between this modified density and the red fraction (for
comparison, Fig.~\ref{fig-baldry:frac-ampl-adj}(a) uses the original density).
Thus, the very local density (the mass/luminosity of the galaxy) and the
neighbor density combine to produce a good predictor of whether a galaxy has
been transformed to the red sequence.

This is consistent with the red distribution being formed from major
interactions (mergers and/or harassment). In a hierarchical formation
scenario, a present-day high-luminosity galaxy will have formed from many
smaller galaxies and therefore is more likely to have undergone a violent
process than a low-luminosity galaxy, regardless of environment.  In other
words, isolated red-distribution galaxies could be regarded as fossil groups
\citep{ponman94,MZ99}. Neighbor density increases the chance of a violent
process either by harassment from other galaxies \citep{MLK98} or by
affecting the impact parameters of merging galaxies/proto-galaxies.

\section{Discussion}
\label{sec-baldry:discussion}

Many of the discussions in the literature have focused on quite specific
mechanisms for different types of galaxies (e.g., see review by
\citet{Fritze04} on S0 galaxies). Here, the analysis shows that there are
essentially two phase spaces for galaxies (in terms of color and absolute
magnitude). Even with the addition of other parameters such as surface
brightness and concentration, the multi-variate distribution is bimodal
\citep{hogg02red,blanton03broadband}; while bimodality is not a ubiquitous
feature of semi-analytic models of galaxy formation
\citep{WF91,kauffmann99,SP99,cole00}. Perhaps the evolution of galaxies is in
some sense simple.  Regardless of the processes (accretion, merging,
harassment, ram-pressure stripping), galaxies will populate one of two regions
in parameter space and while the transformation from the blue to the red
distribution may be deterministic (e.g.\ major merger), the properties within
a distribution depend most strongly on the mass/luminosity of the galaxy and
the effects of environment are mostly chaotic (e.g.\ the effects of minor
mergers depending on impact parameters).

What about morphology? It could be argued that the CM distributions defined
here are reproducing the morphology-density and CM relations of E+S0 and
spiral galaxies. It is not possible to obtain Hubble types for all galaxies in
the SDSS main galaxy sample because the imaging resolution is not sufficient.
Nevertheless, it is unlikely that the red/blue distributions correspond
precisely to these classes. Observers do not always agree on Hubble type even
with many resolution elements and, in the spirit of our interpretation,
whatever processes give rise to the blue/red distribution should also give
rise to {\em distributions} in morphology. Thus, S0 or Sa galaxies could have
a probability of belonging to one or the other distributions and should not be
considered as classes. The method presented in this article could be extended
to include quantified morphological parameters to determine the minor/major
environmental effects in terms of morphology.

What about other populations? This analysis only shows that there are two
dominant populations. There are other distinct populations or distinct
sub-categories. It is important to distinguish between extremes of a
population and populations with separate identities. For example, passive
spirals \citep{couch98,goto03} could be regarded as the extremes of one or
both of the populations, whereas post-starburst galaxies
\citep{zabludoff96,goto03hd} could represent a separate, transforming,
population. In the next section, we discuss a model that includes a
significant fraction of slowly transforming galaxies.

\subsection{Transformation modeling}

One possibility for the trend in population abundance with environment could
be that the mass function varies with local density.  However, observations of
the near-infrared luminosity function show that any change with environment is
likely to be small \citep{depropris98,AP00,balogh01}.  On the other hand, the
strong redshift evolution observed in the colors of galaxies in both clusters
\citep{BO84,fairley02,FZM04} and the field \citep{lilly96,connolly97} suggests
that some galaxies transform from one population to another. In particular,
\citet{bell04red} noted a build up of stellar mass on the red sequence by a
factor of about two between $z\sim1$ and 0, averaged over all environments.

Various mechanisms for these transformations have been proposed.  Some, like
galaxy mergers and ram-pressure stripping by the intracluster medium, occur on
timescales that are short compared with the lifetime of galaxies
\citep{GG72,FN99,moore99}; other processes, such as the gradual starvation of
a system through the removal of gas, result in a slow decline in star
formation \citep{LTC80,BNM00}.

\begin{figure}
\includegraphics[width=0.7\textwidth]{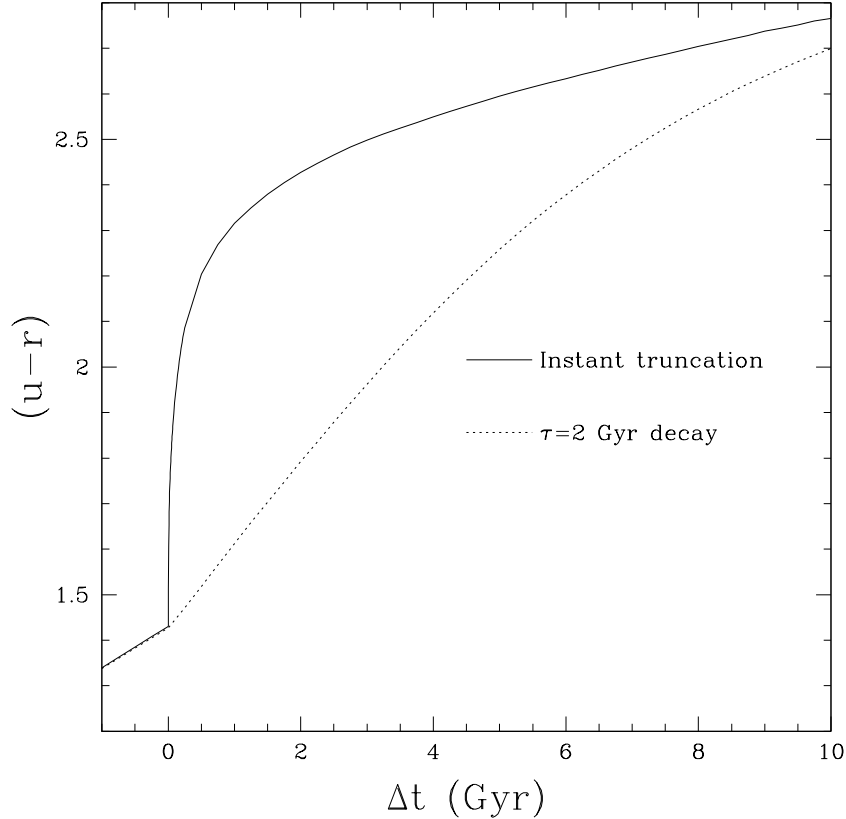}
\caption{The $u-r$ color evolution of two models 
  in which the SFR is reduced at time $\Delta t=0$.  Both models are initially
  evolved for 7\,Gyr ($\Delta t<0$) using a Salpeter IMF and an exponentially
  declining SFR with a timescale $\tau=4$\,Gyr.  The {\it solid line}
  represents a model in which star formation completely ceases at $\Delta
  t=0$, while the {\it dashed line} represents a model in which the SFR
  decreases exponentially with $\tau=2$\,Gyr.}
\label{fig-baldry:modelcomp}
\end{figure}

In Fig.~\ref{fig-baldry:modelcomp} we show the $u-r$ color evolution of two
model galaxies, generated using the {\sc GALEV} population-synthesis code
\citep{BC03}.  We start with a 7\,Gyr old galaxy that has been forming stars
at a rate that has been slowly declining exponentially, with a timescale
$\tau=4$\,Gyr.  Assuming a Salpeter IMF and no dust extinction, this galaxy
has a color $u-r\sim1.4$, which is close to the peak of the observed blue
distribution for the fainter galaxies in our sample.\footnote{The dispersion
  around the color peak could be for a number of reasons: (i) stochastic
  variations in star formation (while the general trend for a population of
  galaxies could be similar, the measured color could vary because of recent
  bursts or quiescent periods); (ii) variations in dust attenuation because of
  disk orientation or intrinsic levels of dust; (iii) variations in
  metallicity; and (iv) photometric errors.} In the first model, we assume all
star formation activity ceases after 7\,Gyr (designated $\Delta t=0$).  In
this case, the galaxy rapidly becomes very red, reaching well within the
observed red distribution ($u-r\sim2$) in less than 0.2\,Gyr.  Therefore, if
galaxies have been transforming at a uniform rate for the last 13.7\,Gyr, we
would only expect to see $<1$\% of them with intermediate colors $1.4<u-r<2$
(in addition to those from the two normal distributions).  This would not have
a noticeable effect on the simplicity of the observed bimodal population.
Thus, it is possible that most or all of the red galaxies have been formed
through short-timescale transformation from the star-forming blue population.
This is consistent with the observed existence of galaxies with short-lived
spectral features indicative of recent changes in star formation history
\citep{DG92,goto03hd,poggianti04,quintero04}.  In this interpretation, the
trend for galaxies to become slightly redder with increasing density is due to
mechanisms (e.g.\ metallicity, dust or previous SFH) that
are independent of this transformation process.

An alternative interpretation is that the transformation is more gradual, as
expected in some models \citep{LTC80,BNM00} and suggested by the lack of
environmental dependence of correlations between different SFH indicators
\citep{kauffmann04}.  The second model in Fig.~\ref{fig-baldry:modelcomp}
shows the color evolution of a galaxy in which, for $\Delta t>0$, the SFR
declines exponentially with a timescale $\tau=2$\,Gyr.  This decline is still
faster than that prior to $\Delta t=0$, but is long enough that the system is
observed with intermediate colors $1.4<u-r<2$ for a substantial amount of time
($\sim 3$\,Gyr).  This could produce a significant distortion on the bimodal
distribution, and might be the cause of the apparent redward shift of the mean
of the blue galaxy distribution with increasing density.

\begin{figure}
\includegraphics[width=1.0\textwidth]{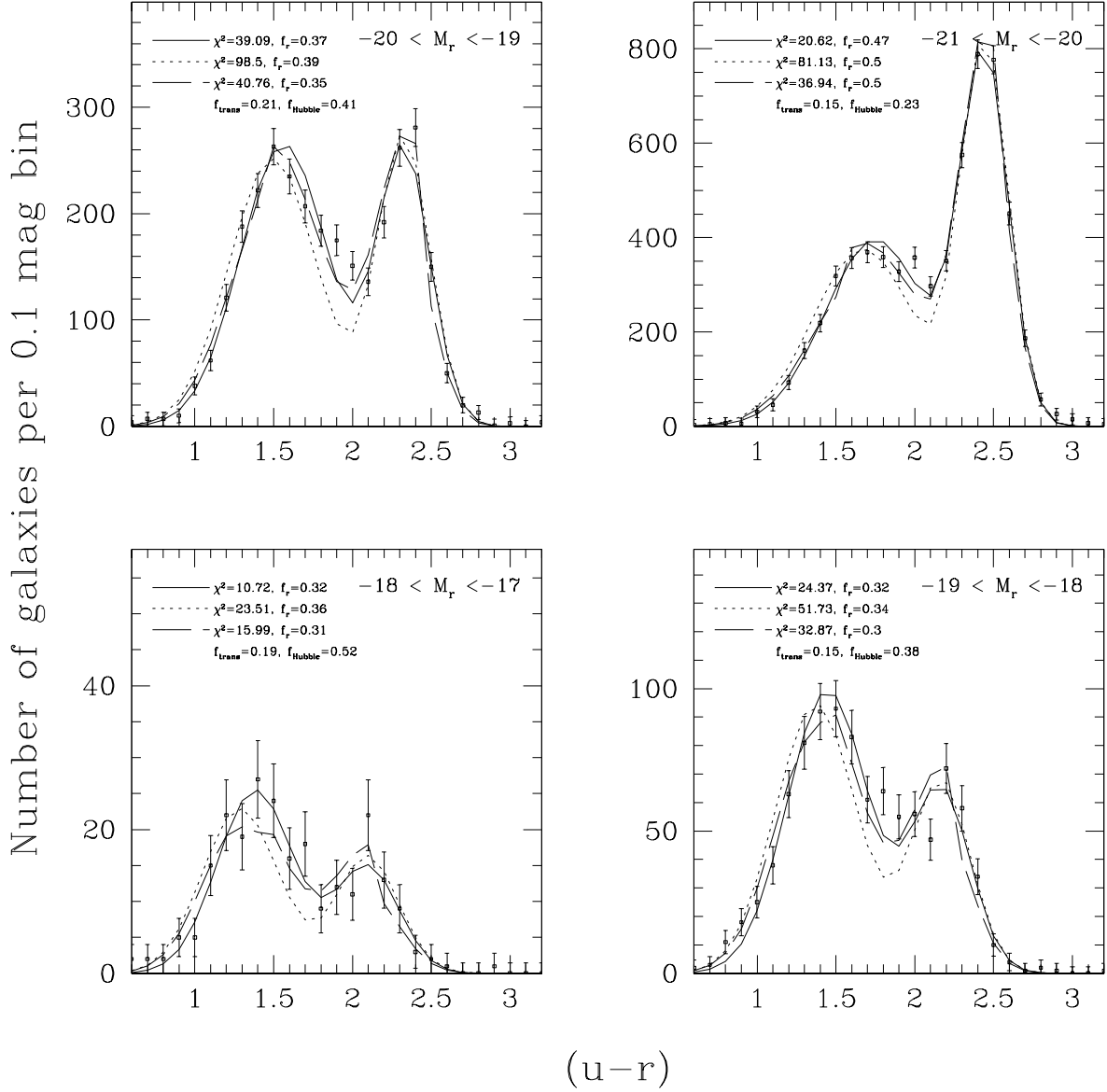}
\caption{Modeling the bimodality. Observed color distributions 
  ({\it points with error bars}) are shown for four luminosity bins,
  restricted to the second-highest density bin ($0.9<\Sigma_5<4.4$).  The {\it
    solid line} shows the double-Gaussian model
  (Fig.~\ref{fig-baldry:color-enviro}), where the mean and amplitude of each
  Gaussian are fit as free parameters.  The {\it dotted line} shows the best
  fit obtained with the mean of the blue distribution constrained to be the
  same as in the fit to the lowest density bin ($\Sigma_5<0.13$).  The {\it
    dashed line} shows the best fit obtained by adding a third component,
  consisting of a population of galaxies with colors intermediate between the
  two peaks, with a distribution given by the rate of color evolution in the
  $\tau=2$\,Gyr model (Fig.~\ref{fig-baldry:modelcomp}).  The $\chi^2$ values
  and the fraction of galaxies in the red distribution ($f_r$) are shown in
  the legend for all of these models.  For the third model, the fraction of
  the total galaxy population comprised by these transforming galaxies at the
  present day ($f_{\rm trans}$) and the fraction of galaxies that would have
  passed through this phase in the last 13.7\,Gyr ($f_{\rm hubble}$), assuming
  the rate of transformation has been uniform, are also shown.}
\label{fig-baldry:transform}
\end{figure}

To demonstrate this, we show in Fig.~\ref{fig-baldry:transform} a series of
model fits to the galaxies in moderately dense environments
($0.9<\Sigma_5<4.4$), where transformations from the star forming
population might be expected to be most common.  We restrict the fits to the
population fainter than $M_r=-21$, where the blue and red populations are most
distinct.  The solid lines show the default, double-Gaussian model fits
presented in Fig.~\ref{fig-baldry:color-enviro}.  The $\chi^2$ value for this
fit, and the fraction of galaxies in the red distribution, are shown in the
legend of each panel.  We now assume that the mean color of the blue
population is independent of density, and equal to the mean that we compute in
the lowest density bin ($\Sigma_5<0.13$).  This model is shown as the dotted
line, and is a poor fit to the data with a much larger $\chi^2$ value in all
luminosity bins.  To model the effect of including a transforming population,
we introduce another parameter, which is the proportion of galaxies in a third
population existing strictly between the means of the blue and red peaks.  The
color distribution of this population over this range is determined by the
amount of time spent at each color, as given by the slope of the dotted line
in Fig.~\ref{fig-baldry:modelcomp}.  For the simple model considered here,
this transforming distribution is nearly uniform in $u-r$ color.  We then fit
the amplitude of this population, as well as those of the red and blue
Gaussian distributions, to minimize the $\chi^2$ of the model. This fit is
shown as the long-dashed line in Fig.~\ref{fig-baldry:transform}.  Although
the $\chi^2$ value is larger than the default two-component model, it still
provides an acceptably good fit.  The fraction of galaxies required to be in
the transforming population is about 15\%--20\%; and when we account for
the duty cycle of this population, up to 50\% of the galaxies may have
undergone such a transformation over the past 13.7\,Gyr, assuming the rate has
remained constant over that time.

We conclude that the best fit to the color distribution in each environment is
provided by the two-population model presented in
refs.~\citep{baldry04,balogh04bimodal}, with the possibility that the entire
population of red galaxies has been built out of transformations from the
bluer population that are ongoing today. These data even accommodate a
relatively slow timescale for this transformation, if it is assumed that the
bulk of the blue galaxy population has a mean color that is independent of
environment. However, a dominance of slow transformations is likely ruled out
by observations of bimodality at $z>1.5$ \citep{somerville04} and by
morphology-color relations, which imply a more violent origin because of bulge
formation.

\clearpage
\begin{theacknowledgments}
  The results presented here made use of the CMU-PITT SDSS Value Added
  Catalog\footnote{\url{http://astrophysics.phys.cmu.edu/dr2_value_added/}}
  created and maintained by K.~Simon Krughoff and Christopher J.~Miller.
  I.\,K.\,B.\ and K.\,G.\ acknowledge generous funding from the David and
  Lucille Packard Foundation.  Funding for the creation and distribution of
  the SDSS Archive has been provided by the Alfred P.\ Sloan Foundation, the
  Participating Institutions, the National Aeronautics and Space
  Administration, the National Science Foundation, the U.S.\ Department of
  Energy, the Japanese Monbukagakusho, and the Max Planck Society.
\end{theacknowledgments}

\end{document}